\documentclass[aps,prl,preprint,groupedaddress]{revtex4-1}
\usepackage{physics}
\usepackage{amssymb}
\def\hf{{\frac{1}{2}}}

\begin{document}


\title{A Dynamic Histories Interpretation of Quantum Theory}


\author{Timothy D. Andersen}
\email[]{andert@gatech.edu}
\affiliation{Georgia Institute of Technology, Atlanta, Georgia}


\date{\today}

\begin{abstract}
The problem of how to interpret quantum mechanics has persisted for a century. The disconnect between the wavefunction state vector and what is observed in experimental apparati has had no shortage of explanations. But all explanations so far fall short of a compelling and complete interpretation. In this letter, I present a novel interpretation called dynamic histories. I show mathematically how quantum mechanics can be reinterpreted as deterministically evolving dynamical world lines in a 5D universe. Quantum probabilities can be then be reinterpreted as stemming from ignorance of the state of our own world line. Meanwhile, the lack of observed superposition in experimental apparati is explained in that we only live on a single history with a definite set of properties. Hence, superposition is not an actual state of a particle but a model of ignorance as in classical probability theory. This explains nonlocal effects without superluminal communication. I also discuss how this relates to 5D Kaluza-Klein theory.
\end{abstract}


\maketitle

\section{Introduction}

Nobel laureate, Steven Weinberg in 2017 published an article \cite{Weinberg2017} called “The Trouble with Quantum Mechanics” in which he laments the lack of a clear interpretation of quantum mechanics. This is, as he says, a debate that has been raging for 100 years with no signs of stopping.

The problem is simply that the current mathematical formalism of quantum mechanics and quantum field theory includes no model or indication of how measurements arise from the predictions of the theory. Rather the theory presents us only with probabilities based on a wavefunction state vector. To quote Erwin Schrödinger \cite{Schrodinger1935}, describing a radioactive decay experiment,

\begin{quotation}
the emerging particle is described … as a spherical wave … that impinges continuously on a surrounding luminescent screen over its full expanse. The screen however does not show a more or less constant uniform surface glow, but rather lights up at one instant at one spot ….
\end{quotation}

There are a handful of interpretations of this phenomenon. These can be classed into complete and incomplete interpretations. Complete interpretations are those that say that the quantum wavefunction is a complete description of a quantum system. Incomplete interpretations are sometimes called hidden variable theories. Albert Einstein favored these with Bohmian mechanics being the simplest.

The problem of quantum interpretation arises from the incompatibility between the basic assumptions of quantum mechanics and the measurement of quantum phenomena. Quantum mechanics makes two basic assumptions:

\begin{enumerate}
	\item The quantum mechanical wavefunction is a complete description of a quantum “microsystem” such as a particle.
	\item Interactions between measurement apparati (including lab equipment and people), i.e. macrosystems, and microsystems are governed completely by Schr\"odinger’s equation. That is, they are linear.
\end{enumerate}

Based on these assumptions Von Neumann gave his idealized general description of measurement (based on the description found in \cite{Bassi2003})

Consider a microscopic system S and select one of its observables O. Let $o_n$ be the eigenvalues of O. Let the spectrum of O be purely discrete and nondegenerate. The corresponding eigenvectors are $\ket{o_n}$. Let M be an apparatus that measures O for the microsystem S. M has a ready state $\ket{A_0}$ where M is ready to measure the property under test and a set of mutually orthogonal states $\ket{A_n}$ that correspond to different macroscopic configurations of the instrument. These are often called pointer positions but they could be any recognizably different measurements.

Now, invoking our assumptions above, the interaction between M and S is linear. We also assume that there is a perfect correlation between the initial state of S and the final state of the apparatus,
$\ket{o_n}\otimes\ket{A_0}$ must yield $\ket{o_n}\otimes\ket{A_n}$. Thus, we are sure that if the state of the apparatus is $\ket{A_n}$, that the state of the particle is $\ket{o_n}$.

The measurement problem is then when we have a superposition of states $\ket{m+l} = 12[\ket{o_m} + \ket{o_l}]$. The linearity of Schr\"odinger’s equation guarantees that the apparatus will be placed into superposition as well: $12[\ket{o_m}\otimes\ket{A_m}+\ket{o_l}\otimes{A_l}]$. The macroscopic measurement apparatus is not in position $m$ or $l$ but a superposition of both. Yet in experiment, macroscopic apparati are always apparently in one or the other state randomly as in a classical mixed state. This leads to the postulate of wavepacket reduction (WPR) that the wavefunction somehow chooses which of its superimposed states to be in.

A more realistic description of measurement that does not assume perfect correlation or a lack of interference from outside sources does not resolve the problem \cite{Bassi2003}. While the interaction between the macroscopic measuring device and the quantum particle undoubtedly disrupts its state in various ways, this disruption is insufficient to explain the paradox.

A resolution to the problem can take three approaches: (1) Violate assumption \#1 and expand beyond the wavefunction to hidden state variables as Bohmian mechanics does, (2) violate assumption \#2 and expand beyond Schr\"odinger’s equation, or (3) keep both assumptions and reinterpret what the wavefunction means.

While incomplete interpretations take the first approach, dynamical reduction models take the second, physically modeling the WPR in the equations.

There is a large body of literature on the 3rd approach. These include superselection theories (both strict and de facto), modal interpretations, consistent histories \cite{Griffiths2003}, and the many worlds and many minds interpretations.

All these approaches have drawbacks. Bohmian mechanics, introduced in 1952 \cite{Bohm1952} and studied continuously since then \cite{Durr2009}, has difficulty with relativity and Lorentz symmetry \cite{Durr2014}, and particle creation and annihilation \cite{Bell1984}. While recent attempts at reconciling with relativity and Bell-type field theories have made progress, these introduce other drawbacks such as randomness \cite{Durr2004}. Bohmian mechanics also cannot eliminate completely its dependence on a universal time parameter, making it difficult to reconcile with general relativity since it requires a foliation \cite{Tumulka2007}. Dynamical reduction models also struggle with extension to relativistic field theories \cite{Bassi2003}.  Superselection \cite{Zurek1982} and modal interpretations \cite{Dieks1998}, meanwhile, suggest that identical wavefunctions in different experiments can have different outcomes. Thus, while they do not have hidden variables, they might as well be incomplete theories. Consistent histories incorporates randomness even though Schr\"odinger's is not a stochastic equation \cite{Griffiths2003}. Many worlds \cite{Dewitt2015} and many minds \cite{Albert1988}\cite{Lockwood1996} are problematic in a variety of ways. Many worlds suggests that observation can change reality itself. Many minds indicates that human consciousness itself splits, somehow needing a classical viewpoint to maintain cohesion, a bold and exotic claim that refutes the universality of quantum mechanics \cite{Dieks1991}. This is hardly an exhaustive list of issues.

In this letter, I present a solution in a 5D universe that suffers from none of these drawbacks. It fits into that rare category that violates both assumptions of quantum mechanics. It does not accept Schr\"odinger's except statistically, and also concludes that the full state of a particle is given by a world sheet description in time and the fifth dimension. I show mathematically that this theory agrees exactly with experiment for both quantum mechanics and relativistic quantum field theory. The principles upon which the theory are based come from classical statistical mechanics and the equivalence of averages over dynamics of systems in equilibrium to averages over configuration space (ergodicity). Since all quantum mechanical and quantum field theoretic systems are essentially averages over configuration space, I show how adding a dynamical component in a fifth dimension (as has been done for 40 years in molecular dynamics simulations) can resolve the quantum interpretation problem.

The resulting theory is deterministic. Randomness arises, not from any source in the universe, but from our own ignorance of our 5th dimensional state.

Finally, I discuss briefly in a non-compactified Kaluza-Klein theory that this dimension can be reconciled with general relativity. Indeed, Kaluza's original theory is the classical limit of this theory. I show why we cannot perceive motion in the 5th dimension as we can in other spatial dimensions because of symmetry breaking. The fifth dimension represents the U(1) phase translation symmetry of electromagnetism which is distinct from the local Lorentz symmetry of the other four.

\section{Equivalence to standard quantum theory}

The equivalence between Schr\"odinger's equation, $i\hbar d\ket{\psi}/dt = \hat{H}\ket{\psi}$, and Feynman's path integral is well known. Suppose $\hat{H}=\hat{p}/2m + V(\hat{q})$ is the Hamiltonian operator and $\psi$ the wavefunction in a Hilbert space, $\hat{p}$ and $\hat{q}$ are the Heisenberg momentum and position observables. The classical Lagrangian functional for a closed system is ${\cal L}(q(t), \dot{q}(t)) = \hf m \dot{q}^2 - V(q)$ where $q$ is the classical position. The classical action is $S=\int_0^T dt\, {\cal L}$. Then the appropriately normalized probability amplitude for a final free-particle spatial state $\ket{F}$ at time $T$ is
$
\braket{F}{\psi(T)} = \mel**{F}{\exp[-\frac{i}{\hbar} \hat{H}T]}{0} = \int Dq(t) \exp\left(\frac{i}{\hbar}S\right)
$
where $\ket{t} \equiv \ket{\psi(t)}$.

Under a Wick rotation $t\rightarrow i\tau$, by analytic continuation, the path integral probability amplitude is equivalent to the partition function for a canonical equilibrium statistical ensemble at temperature $\hbar$, $\int Dq(t) \exp(iS/\hbar) \rightarrow \int Dq(\tau) \exp(-E/\hbar) = Z$ where $E=\int dt H(q(t),\dot{q}(t))$ is the energy integrated over time.

If the probability amplitude exists, then, because it is a continuum integral over Hilbert space, it is equivalent to the microcanonical ensemble up to a constant factor,
\[
Z = C_0\Omega = C_0\int Dq(t) \delta\left(A - E\right),
\] for an appropriate choice of $A$ such that the temperature is $\hbar$. This can be shown by a gradient descent method in a thermodynamic limit, e.g., from a discretized $q$ function to a continuum \cite{Pauli1973}. Microcanonical quantum field theory derives from this equivalence \cite{Strominger:1983}\cite{Strominger:1983b}\cite{Iwazaki:1984}\cite{Iwazaki:1985}.

As in molecular dynamics simulations of quantum lattice gauge theory \cite{Callaway1982}\cite{Callaway1983}, up to a constant factor we add an additional kinetic term to the Lagrangian in a new dimension $y\in \mathbb{Y}$ (generally $\mathbb{Y}=\mathbb{R}$) and add a parameter to $q(t)$ so that it becomes $q(y,t)$. The original classical Hamiltonian becomes the potential energy and the new Lagrangian is $\bar{\cal L}(q_y,\dot{q},q) = q_y^2/2 - H(\dot{q},q)$. The new action is $\bar{S}=\int dy dt\, \bar{\cal L}$. It is easy to show that both the canonical and microcanonical ensembles in this action, e.g., $\bar{Z}=\int Dq_y Dq \exp(-\bar{S}/\hbar)$, are equivalent up to a constant $\bar{Z}=C_1Z$. The kinetic term can simply be integrated out of the functional integral since it is quadratic.

Now, we can apply Euler-Lagrange to the new action $\bar{S}$ to obtain equations of motion in $y$,
\begin{equation}
q_{yy} = \frac{\partial\bar{\cal L}}{\partial q} - \frac{\partial}{\partial t}\frac{\partial \bar{\cal L}}{\partial \dot{q}}
\label{eqn:el}
\end{equation} The right hand side are the classical equations. If we solve these for $q$, we obtain a solution $\bar{q}(y,t)$, given appropriate stochastic initial conditions $\bar{q}(y,0)$ that satisfy the Born rule and the given lab preparation. (Thus, while the equations are not stochastic, the  initial condition makes it impossible to predict. This reflects our ignorance of the exact state of $q(y,0)$ at any $y$.) 

For example, in the free particle case, the equations of motion (in real time) are the one dimensional wave equation: $q_{yy} = q_{tt}/m$, the so-called guitar string equation. This solution involves two pulses (or moving waves) the right and the left that move forward and backward in time respectively as they move in the $y$ direction. The velocity in $y$ is inversely proportional to the square root of the mass of the particle (and the speed of light which here is unity), suggesting that macroscopic objects change their position little in the free case between world lines.

By ergodicity, we can obtain the expected value of any classical observable $O(q(y,t))$ by averaging it over all $y$, $\expval{O(q(t))} = \lim_{Y\rightarrow\infty} \frac{1}{Y}\int_0^Y dy O(\bar{q}(y,t))$. Thus, for a final configuration $F$, one need only apply the correct observable for the measurement.

Analytic continuation of any predictions back to real time can be done at any point.

As I showed in a previous paper, this can be done as well for all of quantum field theory in 5D. This has been done since the early 1980s in lattice simulations for QED, electroweak, and QCD using a Hamiltonian-based molecular dynamics approach \cite{Andersen2019}. (Modern implementations are typically hybrid Monte Carlo which includes an acceptance/rejection step, but the Hamiltonian dynamical principle is still used.) My paper simply extends this to perturbation theory of continuous fields showing how vacuum loops can arise from the interaction of various particle world lines in the fifth dimension.

\section{How to interpret the equations of motion}
The key step in interpreting quantum mechanics is understanding the leap from the equations of motion to generating averages of observables or probability amplitudes. While the statistical averaging is equivalent to Schr\"odinger's equation, the equations of motion contain much more information. Likewise, while the wavefunction $\psi(t)$ contains sufficient information to compute probabilities, the state is determined by the dynamics of $q(y,t)$. I.e., it is not a probability amplitude of world lines $q(t)$ but an evolving world sheet $q(y,t)$. Of course, $q$ need not be position. It could be any variable or field of interest.

It is clear that if what is measured in our measurement scheme is not a superposition but a single result or set of results, then what we are measuring is not some average over $q(y,t)$ but a particular $q(y_F,t)$. In other words, the equations of motion in 5D give us a particular world line to measure, but we are not able to determine, because of ignorance of the boundary conditions, what the full state is. Instead, we can only access a property of $y_F$ which tells us only that $y_F$ is from the subset that contains worldlines with that property.

With this in mind, let us revisit the Von Neumann measurement scheme. Given the superposition of states we know that a subset of $\mathbb{M}\subset\mathbb{Y}$ have property $\ket{o_m}$ while another equal subset $\mathbb{L}\subset\mathbb{Y}$ have property $\ket{o_l}$. For any given $y\in\mathbb{Y}$, a particle on a world line $q(y,t)$ can have only one property.

Upon carrying out an experiment, both particle, experimenter, and apparatus share a single position in $y$ as they share a position in time $t$. The fifth dimension is assumed to be orthogonal to time. Thus, the particle world line as well as the apparatus' worldline are in motion in $y$ forming world sheets $q(y,t)$ and $Q(y,t)$. The particle state $\ket{m+l}$ represents all possible worldlines in $\mathbb{Y}$. Yet, the particle, by equations of motion \ref{eqn:el} only takes on a single world line with a single property. Thus, the superposition of states represents our ignorance of which world line is the correct one, not an intrinsic property of the particle.

The experimental measurement is made at a particular $y_F$ and $t_F$ where the particle terminates. (This is an ideal statement since measurement takes some non-zero amount of time as well to decohere the particle wavefunction.) The apparatus now reflects the property $\ket{o_m}$ or $\ket{o_l}$ that the wavefunction on $y_F$ has. By gaining this knowledge, we now reduce our ignorance of which subset of $\mathbb{Y}$ we are on, e.g., given property $\ket{o_m}$ we know we are in subset $\mathbb{M}$.

Since the measurement occurs at a particular $y_F\in\mathbb{Y}$ it must fall into either subset $\mathbb{M}$ or $\mathbb{L}$ exclusively. Because we cannot predict from the outset, however, into which it falls, it appears random. 

Once the measurement is made, as in the consistent histories interpretation \cite{Griffiths2003}, it appears as if the particle had that property all along even though this is not true.

Note that the dynamics in $y$ is for all of history, not just at the present time. The equations are similar to those of a moving string that propagates through time. While we can only perceive history as belonging to the subsets of $\mathbb{Y}$ for which we have made measurements, history itself belongs to one and only one $y$ not a combination of different $y$ at different $t$, which could lead to historical inconsistencies.

This approach resolves quantum paradoxes as arising from ignorance and increasing knowledge about our worldline. For example, in the case of Schr\"odinger's cat, some world lines include a dead cat while others include a live cat. Thus, if the experiment is conducted at time $t$, the cat's world sheet $C(y,t)$ will contain a subset $\mathbb{D}\subset\mathbb{Y}$ and a subset $\mathbb{A}\subset\mathbb{Y}$ for the dead and alive cats. Yet, if our world line is at $y_F$, then it will determine the cat's ultimate fate.

The Einstein-Podolsky-Rosen paradox \cite{Einstein1935}, which exposes the nonlocal nature of quantum mechanics, can also be dealt with without superluminal communication. In this case, using David Bohm's version \cite{Bohm1960}\cite{Bell1964}, an electron positron pair is produced as a spin singlet. Alice measures the electron spin while Bob measures the positron spin. Depending on the set up of the measurement apparatus, Bob's measurement is 100\% correlated with Alice's or not at all. If Alice measures the property $+z$ she knows that her world line, $y_A$, is in a subset $\mathbb{S}_z\subset\mathbb{Y}$ that has that property. When Bob makes his measurement, he can choose to measure the z-axis as well, applying $S_z$ to the wavefunction. What this does is selects out the subset of worldlines from $\mathbb{Y}$ that have the z-property that Bob measures. When he compares notes with Alice, he and Alice are on the same world line. Therefore, their world line must have consistent z-spin properties. Hence they know their world line is in a subset $\mathbb{S}_z\subset\mathbb{Y}$ from Alice's measurement alone. Bob contributes no new information. 

If Bob chooses to measure the x-spin propety, however, Alice is ignorant of this property; thus, Bob has a random chance of measuring $+x$ or $-x$. When he compares notes with Alice, he will find that between the two of them they have reduced their knowledge of the subset of $\mathbb{Y}$ to be that which has the property Alice measured and the one he measured, $\mathbb{S}_z\cap\mathbb{S}_x\subset \mathbb{Y}$. Thus, in this case, their ignorance is further reduced. 

Alice, however, cannot gain this knowledge by measuring the electron by itself because of the non-commutative nature of $S_z$ and $S_x$. 

Non-commutativity is a feature of stochastic processes in general such as Brownian motion \cite{Kauffman2004} as Feynman noted in his development of the path integral \cite{Feynman1948}. In the dynamical case, it is guaranteed by Alice and Bob's uncertainty about their world line state, which causes state knowledge to diffuse (in imaginary time) stochastically.

What this shows is that measurement is a process of applying observable operators to reduce ignorance about what subset of $\mathbb{Y}$ we are living on. Randomness is not inherent in the universe, or, as Einstein would say, God (or Nature) does not play dice. Randomness is guaranteed only by our inability to know the full state of our world line.

\section{How the universe can appear 4D}
While the issue of quantum field theory has been largely addressed in my previous paper \cite{Andersen2019}, the problem of how this fits in with the 4D general relativity is still an issue. In particular, how does the fifth dimension's symmetry group appear in quantum theory? Kaluza-Klein theory posits that the universe is 5D and shows how the fifth dimension plays the role of U(1) symmetry of electromagnetism. I will not go into the mathematical formalism of non-compactified KK theory; see \cite{Overduin:1997} for a detailed discussion. I will also not address how other forces or matter fit into the theory. The problems with going beyond 5D theories are substantial \cite{Witten1981}. I will only address the direct implication of adding a dimension to spacetime.

Kaluza's original theory proposed that the universe was constant in the 5th dimension. While this is the classical limit of the quantum theory presented here, in reality there would have to be variations at all scales in the 5th dimension to account for quantum effects, at least down to the Planck scale. Compatification, however, appears to be unnecessary here.

One of the main reasons the fifth dimension is not navigable as spatial dimensions are, even if it is spacelike as in many KK theories, is because of the symmetry breaking between gravity and electromagnetism in the KK theory. While local Lorentz boosts and rotations are possible in four of the dimensions, in the fifth dimension de Sitter rotations and boosts become phase translations in electromagnetic fields. Thus the local de Sitter symmetry, SO(4,1), has been broken into local Lorentz symmetry, SO(3,1), and U(1).

Even if the fifth dimension is timelike, given the statistical results from quantum theory, entropy is constant and maximal in that dimension, i.e., it is in equilibrium. In a constant entropy, equilibrium system there is no way to distinguish which points are past and which are future \cite{Maccone2009} and no way to do work because there is no way to flow free energy in a single direction. There is no arrow. 

It is impossible even in principle to measure motion through a maximally entropic dimension without doing work in another dimension that is not in equilibrium such as time. A good example of this is that one can create an equilibrium Brownian clock by measuring the distance a particle travels to be $\propto\sqrt{t}$, but one cannot record this information in the Brownian system itself.

\section{Conclusion}
I have shown how quantum mechanics is equivalent to averaging over world sheets in a 5th dimension orthogonal to the ordinary four. In doing so, I have demonstrated that measurement is a process of applying observable operators to state vectors to reduce our ignorance of what world line we live on within that dimension. This explains the process of quantum measurement without wavefunction collapse (only our ignorance collapses). The theory is completely deterministic. It also does not invoke many worlds or minds. It allows for many different world histories to exist and for the past and future to change in the fifth dimension. Yet, only one history ever needs to exist at a ``time'' since history itself is dynamical. Unlike Many Worlds, there is no splitting when observations are made. (What splits is our knowledge of what subset we live on). One startling conclusion however is that in the dynamical history theory history itself is not fixed. Rather history from the Big Bang to the end of time can change all at once as it propagates in the 5th dimension. We would have no knowledge of this change since history always remains consistently on a single world line.

\bibliography{q5d}

\end{document}